\documentclass[11pt]{article}
\usepackage{amsmath,amsfonts,amssymb,latexsym}

\def\be{\begin{equation}}
\def\ee{\end{equation}}
\def\bq{\begin{eqnarray}}
\def\eq{\end{eqnarray}}
\def\beq{\begin{eqnarray*}}
\def\eeq{\end{eqnarray*}}

\begin{document}

\title{{\Huge Series expansions and sudden singularities}}
\author{{\Large John D. Barrow$^{1}$\footnote{\texttt{email:jdb34@hermes.cam.ac.uk}}\,,\,\,  
S. Cotsakis$^{2}$\footnote{\texttt{email:skot@aegean.gr}}\,\, and 
A. Tsokaros$^{2}$\footnote{\texttt{email:atsok@aegean.gr}}} \\
$^{1}$DAMTP, Centre for Mathematical Sciences, \\
Cambridge University, Cambridge CB3 0WA, UK \\
$^{2}$ GEO.DY.SY.C. Research Group,  \\ University of the Aegean, Karlovassi 83200, Samos, Greece}
\maketitle

\begin{abstract}
\noindent We construct solutions of the Friedmann equations near a sudden
singularity using generalized series expansions for the scale factor, the
density, and the pressure of the fluid content. In this way, we are able to
arrive at a solution with a sudden singularity containing two free
constants, as required for a general solution of the cosmological equations.
\end{abstract}


\noindent A sudden singularity will be said to arise everywhere at comoving
proper time $t_{s}$ in a Friedmann universe expanding with scale factor $a(t)
$ if
\begin{equation}
\lim_{t\rightarrow t_{s}}a(t)=a_{s}\neq 0,\qquad \lim_{t\rightarrow t_{s}}%
\dot{a}(t)=\dot{a}_{s}<\infty ,\qquad \lim_{t\rightarrow t_{s}}\ddot{a}%
(t)=\infty ,  \label{def}
\end{equation}%
for some $t_{s}>0$, and we have set $a(t_{s})\equiv a_{s}$. In standard
Friedmann-Lema\^{\i}tre cosmological models a fluid is placed in a
homogeneous and isotropic spatial geometry whose dynamics is then determined
by two independent Einstein equations for three unknown time-dependent
functions, the Friedmann metric scale factor, $a(t)$, the fluid density, $%
\rho (t)$, and fluid pressure, $P(t)$, respectively (with units chosen with $%
c=8\pi G=1$):
\begin{equation}
\frac{\dot{a}^{2}+k}{a^{2}}=\frac{\rho }{3},\quad -2\frac{\ddot{a}}{a}-\frac{%
\dot{a}^{2}+k}{a^{2}}=P.  \label{frw}
\end{equation}%
When an equation of state $P=f(\rho )$ is chosen this system closes, but we
look for solutions of these equations having a sudden singularity, when is
no equation of state connecting $P$ and $\rho $.

From the Definition (\ref{def}), it follows that Taylor series are obviously
not general enough to describe the local behaviour of solutions near a
sudden singularity, and in \cite{scot,bct1,bct2} we began the analysis of
such solutions in terms of various types of generalized power series. We use
the term \emph{Puiseux series} for an expansion of the form
\begin{equation}
x(t)=\sum_{i=0}^{\infty }a_{i}\ (t_{s}-t)^{i/s};
\end{equation}%
that is, with rational exponents and a constant first term, and we refer to
a series development of the form
\begin{equation}
x(t)=(t_{s}-t)^{h}\sum_{i=0}^{\infty }a_{i}\ (t_{s}-t)^{i/s},
\end{equation}
as a \emph{Fuchs series}; that is, when there is an additional real indicial
exponent $h$ and no constant term (thus Fuchs and Puiseux series here
correspond to Frobenius and Taylor series respectively when $s=1$, cf. \cite%
{bender}). Setting $x=a,\ y=\dot{a},$ we write the original equations (\ref%
{frw}) in first-order form,
\begin{equation}
\dot{x}=y,\quad \quad \dot{y}=-\frac{k+y^{2}+x^{2}P}{2x}.  \label{eq:DynSys}
\end{equation}%
Below we show that in a dominant solution of this system near a
sudden singularity, with leading behaviour $x=a_{s}$ and $y=\dot{a}_{s}$
both remaining finite but with the pressure being infinite, all variables of
the problem can be expressed in terms of generalized power series with the
above forms. In particular, we can show that all solutions of (\ref%
{eq:DynSys}) must have the form of Puiseux series
\begin{equation}
x(t)=\sum_{i=0}^{\infty }c_{1i}\ (t_{s}-t)^{i/s},\quad
y(t)=\sum_{i=0}^{\infty }c_{2i}\ (t_{s}-t)^{i/s},  \label{eq:DynSysSol}
\end{equation}%
where $s>1$, and
\begin{equation}
c_{10}=\alpha \neq 0,\quad c_{20}=\beta \neq 0,
\end{equation}%
whereas the pressure must be of the form of a Fuchs series,
\begin{equation}
P=(t_{s}-t)^{h}\sum_{i=0}^{\infty }p_{i}\ (t_{s}-t)^{i/s},
\label{eq:PresSer}
\end{equation}%
with the restriction that $h<0$.

By substituting these forms into the field equations, and balancing the
various terms, we can determine the series expansions of the scale factor,
density and pressure that will lead to a specific sudden singularity. For
example, if we choose $n=3/2$ in the original solution with a sudden
singularity of the form \cite{jdb04a},
\begin{equation}
a(t)=\left( \frac{t}{t_{s}}\right) ^{q}(a_{s}-1)+1-\left( 1-\frac{t}{t_{s}}%
\right) ^{n},
\end{equation}%
then we find that
\begin{eqnarray}
a(t) &=&\left( \frac{t}{t_{s}}\right) ^{q}(a_{s}-1)+1-\left( 1-\frac{t}{t_{s}%
}\right) ^{3/2}=1-\frac{1}{t_{s}^{n}}(t_{s}-t)^{3/2}+\sum_{i=0}^{\infty
}a_{i}(t_{s}-t)^{i} \\
&=&(1+a_{0})+a_{1}(t_{s}-t)-\frac{1}{t_{s}^{3/2}}%
(t_{s}-t)^{3/2}+a_{2}(t_{s}-t)^{2}+\cdots =\sum_{i=0}^{\infty
}b_{1i}(t_{s}-t)^{i/2},
\end{eqnarray}%
where
\begin{equation}
b_{10}=a_{s},\,b_{11}=0,\,b_{12}=a_{1},\,b_{13}=-\frac{1}{t_{s}^{3/2}},\,%
\mbox{and for\, }i\geq 4\,b_{1i}=\left\{
\begin{array}{cc}
0 & i\mbox{ odd,} \\
a_{i/2} & i\mbox{ even.}%
\end{array}%
\right.
\end{equation}%
In this case, the corresponding asymptotic expansions for the pressure and
density are given by the following forms (these were first constructed in
Ref. \cite{bct1}):
\begin{equation}
P=\frac{3}{2\alpha t_{s}^{3/2}}(t_{s}-t)^{-1/2}-\frac{4\alpha a_{2}+\beta
^{2}+k}{\alpha ^{2}}-\frac{9\beta }{2\alpha ^{2}t_{s}^{3/2}}%
(t_{s}-t)^{1/2}+\cdots ,
\end{equation}%
with
\begin{equation}
a_{i}=(a_{s}-1)\binom{q}{i}\left( -\frac{1}{t_{s}}\right) ^{i},
\end{equation}%
and there are \emph{two} independent constants, namely,
\begin{equation}
\alpha =1+a_{0}=\alpha _{s},\,\text{(or essentially}\,t_{s})\,\text{and}%
\,\beta =-a_{1},
\end{equation}%
The density has the form
\begin{equation}
\rho =\rho _{s}+\frac{9\beta }{\alpha ^{2}t_{s}^{3/2}}(t_{s}-t)^{1/2}+\cdots
,\,\text{with}\,\,\rho _{s}=\frac{3(k+\beta ^{2})}{\alpha ^{2}}\ .
\end{equation}%
Conversely, starting from the Fuchs expansion
\begin{equation}
P=\sum_{i=0}^{\infty }p_{i}\ (t_{s}-t)^{\frac{i}{2}-\frac{1}{2}},
\end{equation}%
with
\begin{equation}
p_{0}=\frac{3}{2\alpha t_{s}^{3/2}},\qquad p_{1}=-\frac{4\alpha a_{2}+\beta
^{2}+k}{\alpha ^{2}},\qquad p_{2}=-\frac{9\beta }{2\alpha ^{2}t_{s}^{3/2}},
\end{equation}%
and $\alpha ,\beta ,a_{i}$ as above, the system has the following solution
for the scale factor,
\begin{equation}
a(t)=\alpha -\beta (t_{s}-t)-\frac{2\alpha p_{0}}{3}(t_{s}-t)^{3/2}-\frac{%
\alpha ^{2}p_{1}+\beta ^{2}+k}{4\alpha }(t_{s}-t)^{2}-\frac{6\beta
p_{0}+2\alpha p_{2}}{15}(t_{s}-t)^{5/2}+\cdots ,
\end{equation}%
and the next term will contain $p_{0},p_{1},p_{3},$ and so on; that is, we
find the original form,
\begin{equation}
a(t)=\left( \frac{t}{t_{s}}\right) ^{q}(a_{s}-1)+1-\left( 1-\frac{t}{t_{s}}%
\right) ^{3/2}.
\end{equation}%
If in the Fuchs series for the pressure we use a different value of the
indicial exponent $h$, for instance, and we have a total exponent of the
form $\frac{i}{2}-1$, or $\frac{i}{2}-\frac{3}{2}$, then the results are
essentially the same and the density will be given by:
\begin{equation}
\rho =3\frac{k+\dot{a}^{2}(t)}{a^{2}(t)}.
\end{equation}%
As another example, if
\begin{equation}
P=\sum_{i=0}^{\infty }p_{i}\ (t_{s}-t)^{\frac{i}{3}-\frac{2}{3}},
\end{equation}%
then we find
\begin{equation}
a(t)=\alpha -\beta (t_{s}-t)-\frac{9\alpha p_{0}}{8}(t_{s}-t)^{4/3}-\frac{%
9\alpha p_{1}}{20}(t_{s}-t)^{5/3}+\cdots ,
\end{equation}%
with $\alpha ,\beta $ both free constants.

We therefore conclude that with the use of the generalized series methods
considered in this paper, the original, 1-parameter solution with a sudden
singularity of Ref. \cite{jdb04a}, becomes part of a two-parameter, general
solution of the field equations in the isotropic and homogeneous case with a
sudden singularity. A similar result for the general inhomogeneous case has
been previously reached in \cite{bct2}, thus enhancing the viability of
solutions with sudden singularities.

\end{document}